\documentclass{iopart}
\usepackage{epsfig,newlfont}
\DeclareMathAlphabet{\mathitbf}{T1}{cmr}{bx}{it}
\usepackage[colorlinks, 
linkcolor=blue,
urlcolor=blue,
citecolor=blue,
bookmarksopen=false]{hyperref}

\newcommand{\romanb}{{\mathrm{b}}}
\newcommand{\romanc}{{\mathrm{c}}}
\newcommand{\romand}{{\mathrm{d}}}
\newcommand{\romang}{{\mathrm{g}}}
\newcommand{\romane}{{\mathrm{e}}}
\newcommand{\romanh}{{\mathrm{h}}}
\newcommand{\romani}{{\mathrm{i}}}
\newcommand{\romanj}{{\mathrm{j}}}
\newcommand{\romanr}{{\mathrm{r}}}
\newcommand{\romant}{{\mathrm{t}}}
\newcommand{\romanM}{{\mathrm{M}}}

\begin{document}
 
\title[Bootstrap current at TJ-II]{Calculation of the bootstrap current
profile for the TJ-II stellarator}

\author{J.L. Velasco$^1$, K. Allmaier$^2$, A. L\'opez-Fraguas$^1$,
C.D. Beidler$^3$, H. Maassberg$^3$, W. Kernbichler$^2$,
F. Castej\'on$^1$, J.A. Jim\'enez$^1$}

\address{$^1$ Laboratorio Nacional de Fusi\'on, Asociaci\'on EURATOM-CIEMAT, Madrid, Spain
\\ $^2$ Institut f\"ur Theoretische Physik - Computational Physics, Technische Universit\"at Graz, Association EURATOM-\"OAW, Graz, Austria
\\ $^3$ Max-Planck Institut f\"ur Plasmaphysik, IPP-EURATOM, Greifswald, Germany}

\ead{joseluis.velasco@ciemat.es}

\begin{abstract}

  Calculations of the bootstrap current for the TJ-II stellarator are
  presented. DKES and NEO-MC codes are employed; the latter has
  allowed, for the first time, the precise computation of the
  bootstrap transport coefficient in the long mean free path regime of
  this device. The low error bars allow a precise convolution of the
  monoenergetic coefficients, which is confirmed by error
  analysis. The radial profile of the bootstrap current is presented
  for the first time for the 100\_44\_64 configuration of TJ-II for
  three different collisionality regimes. The bootstrap coefficient is
  then compared to that of other configurations of TJ-II regularly
  operated.  The results show qualitative agreement with toroidal
  current measurements; precise comparison with real discharges is
  ongoing.

\end{abstract}


\section{Introduction}\label{SEC_INTRO}

The bootstrap current is a neoclassical effect triggered by the radial
gradients of the density and temperature in the presence of an
inhomogeneous magnetic field, see
e.g. Refs.~\cite{peeters2000bootstrap,helander2002collisional} and references therein.

The control of the bootstrap current, and thus of the total parallel
current, may lead to the possibility of continuous operation in
tokamak overdense plasmas. In stellarators, it can provide access to
improved confinement regimes, by means of control of the rotational
transform profile. On the other hand, the bootstrap current may
perturb the desired magnetic configuration produced by the external
coils. This is specially important for shearless devices, such as W7-X~\cite{turkin2006current}
or TJ-II~\cite{alejaldre1999first}.

Indeed, one of the main lines of research at the flexible heliac TJ-II
is the relation between confinement and the magnetic
configuration~\cite{alejaldre1999confinement,ascasibar2002confinement,ascasibar2005magnetic,castellano2002well,lopez2009cmode}. Ref.~\cite{castejon2005topology}
offers a review of results at TJ-II and other stellarators. Therefore,
an estimation of finite-$\beta$ effects on the characteristics of the
magnetic configuration such as the rotational transform, the magnetic
shear, the position of rational surfaces or the magnetic well, is of
great importance. These effects are not expected to be large, since
TJ-II is a heliac-type stellarator~\cite{alejaldre1999first}
consequently showing a small Shafranov shift of the magnetic surfaces,
but still this should be quantified.

Additionally, an Electron Bernstein Wave (EBW) heating system is being
installed at TJ-II~\cite{fernandez2005ebw}. Apart from its main
purpose of heating overdense plasmas, it may be employed to drive
current as well. Then, the total current (bootstrap plus the one driven by
EBW) could be used to try to tailor the rotational transform
profile~\cite{garcia2011ebw,castejon2011ebwbc}.  \\

The calculation of the bootstrap current is a numerical challenge for
non-axisymmetric devices, since the error estimates for computations
in the long-mean-free-path ({\em lmfp}) regime of stellarators are
usually large: DKES~\cite{hirshman1986dkes} has a convergence problem
in the {\em lmfp} and the Monte Carlo methods have an unfavorable
scaling of the computing time with the collisionality. Examples of
similar calculations for non-axisymmetric devices may be found in
Refs.~\cite{maassberg2005ecd,isaev2009bootstrap}. The dependence of
the bootstrap current on the configuration has been explored in
several devices such as W7-X~\cite{maassberg2009momentum} and
Heliotron-J~\cite{nishimura2009viscosity}. The former presents
momentum correction techniques applied to the calculation of the
bootstrap current profile for two different configurations of W7-X. In
the latter, the monoenergetic bootstrap coefficient is studied as a
function of the {\em bumpiness} of the
configuration. Ref.~\cite{beidler2011ICNTS} also contains a
comprehensive study of the numerical results for the bootstrap
coefficient, as well as the other monoenergetic transport
coefficients, for several stellarators.

Despite its interest, the bootstrap current profile at TJ-II is not
known with precision: the problem of lack of accuracy in the {\em
lmfp} regime is especially relevant for this
device~\cite{ribeiro1987plateau,yunta1990bootstrap,tribaldos2003bootstrap}, which is
characterized by a very complex magnetic configuration. Recently, some
effort has been put in calculating with precision the bootstrap
coefficient, see Ref.~\cite{beidler2011ICNTS} and references therein. One of
the codes in Ref.~\cite{beidler2011ICNTS} is NEO-MC~\cite{allmaier2008variance}, which
was explicitly developed in order to overcome the problem of poor
convergence in the {\em lmfp} regime. It combines the standard $\delta
f$ method (see Ref.~\cite{allmaier2008variance} and references therein) with an
algorithm employing constant particle weights and re-discretizations
of the test particle distribution. Apart from the monoenergetic
bootstrap coefficient, NEO-MC also calculates the radial diffusion
coefficient, required for calculating the ambipolar electric field,
and the parallel conductivity coefficient, required for the momentum
correction techniques~\cite{maassberg2009momentum,sugama2002viscosity}.

In this work, we show that NEO-MC is able to provide, for the first
time, calculations of the contribution of the {\em lmfp} regime to the
bootstrap current of TJ-II with high accuracy. We present, for the
100\_44\_64 magnetic configuration of TJ-II, computations of NEO-MC of
the monoenergetic bootstrap coefficient at several radial positions
for a wide range of collisionalities and radial electric fields. This
allows, by means of energy convolution and momentum correction
technique, to estimate the profile of the bootstrap current. A
Monte-Carlo technique for calculating error propagation allows us to
state quantitatively that the results shown are of significance.

TJ-II is a flexible Heliac whose magnetic configuration can be
modified, both on a shot-to-shot
basis~\cite{alejaldre1999confinement,ascasibar2002confinement,ascasibar2005magnetic,castellano2002well}
or continuously~\cite{lopez2009cmode}. The bootstrap current may change,
even qualitatively, for different magnetic configurations of
TJ-II~\cite{yunta1990bootstrap}. We have thus explored part of the set of magnetic
configurations of TJ-II with DKES calculations. 
\\

This work is organized as follows. In Section~\ref{SEC_CALCULATION},
we describe the calculation of the bootstrap current: we present the
equations, we outline the difficulties that may arise when solving
them and we briefly describe the algorithm employed. Then, we
particularize the discussion to the TJ-II stellarator: in
Section~\ref{SEC_RESULTS}, we show the results. We first discuss the
monoenergetic calculations, then the profile of the current and
finally the scan in magnetic configurations.
Section~\ref{SEC_CONCLUSIONS} is devoted to the conclusions.  \\

\section{Basic theory and calculation method}\label{SEC_CALCULATION}

In order to estimate the bootstrap current, one has to solve the Drift
Kinetic Equation (DKE) that describes the evolution of the particle
distribution function~\cite{hirshman1986dkes}:
\begin{equation}
\left( \frac{\partial}{\partial t} +
\mathbf{v}_\romang\cdot\mathbf{\nabla} - \cal L_\romanc\right) \hat
f\!=\!\mathbf{v}_\romang\cdot\mathbf{\nabla}\psi\,,
\label{EQ_DKE}
\end{equation}
where $\mathbf{v}_\romang$ is the guiding center drift velocity,
$\psi$ is the toroidal magnetic flux through the local
magnetic surface and $\cal L_\romanc$ is the collision
operator. Here $\hat f$ is the normalized perturbation of the distribution
function $f$ and it is defined through the local Maxwellian
distribution function $f_\romanM$:
\begin{equation}
f\!=\!f_\romanM - \hat f \frac{\partial f_\romanM}{\partial\psi}\,.
\end{equation}

The neoclassical ordering~\cite{helander2002collisional} allows to neglect the
time-dependence of $\hat f$, as well as its radial variation. The
equations become {\em local}, i.e., the radial position enters only as
a parameter. If the $\vec E\times\vec B$ drift included in $\vec
v_\romang$ is treated as incompressible~\cite{beidler2007icnts}, the
neglect of energy diffusion in the Lorentz collision operator allows
for monoenergetic solution of the equations: the kinetic energy can
also be considered a parameter. The remaining equation is then
three-dimensional.

Ignoring momentum-correction for the moment, the fluxes may be
expressed as linear combinations of the gradients of density and
temperature and of the electric field. The following expressions,
together with their derivation, may be found for instance in
Ref.~\cite{beidler2011ICNTS}. For each species $\romanb$, the radial
particle flux and the bootstrap current are:
\begin{eqnarray}
 \frac{\langle\Gamma_\romanb\rangle}{n_\romanb}\!=\!
  - L_{11}^\romanb\left(\frac{1}{n}\frac{\romand n}{\romand r} - Z_\romanb e\frac{E_\romanr}{T_\romanb}-\frac{3}{2}\frac{1}{T_\romanb}\frac{\romand T_\romanb}{\romand r}\right) - L_{12}^b\frac{1}{T_\romanb}\frac{\romand T_\romanb}{\romand r}\,,\label{EQ_LINEAR1}\\
  \frac{\langle\vec j_\romanb\cdot\vec B\rangle}{Z_\romanb e nB_0}\!=\!
  - L_{31}^\romanb\left(\frac{1}{n}\frac{\romand n}{\romand r} - Z_\romanb e\frac{E_\romanr}{T_\romanb}-\frac{3}{2}\frac{1}{T_\romanb}\frac{\romand T_\romanb}{\romand r}\right) - L_{32}^b\frac{1}{T_\romanb}\frac{\romand T_\romanb}{\romand r}\,.
\label{EQ_LINEAR2}
\end{eqnarray}
As usual, $Z_\romanb$, $T_\romanb$ and $n$ are the charge number,
temperature and density of species $\romanb$, $e$ is the elemental
charge and $E_\romanr$ is the radial electric field;
$\romanb\!=\!\romane$ are electrons and $\romanb\!=\!\romani$ are
protons. $B_0$ is the (0,0) Fourier harmonic of the field in Boozer
coordinates and the brackets denote flux-surface average. The thermal
coefficients $L_{\romani\romanj}$ at each radial position are
calculated by convolution with a Maxwellian distribution of the
monoenergetic coefficients:
\begin{equation}
  L^\romanb_{\romani\romanj}(r,n,T_\romani,T_\romane,E_\romanr)\!=\!\frac{2}{\sqrt{\pi}}\int_0^\infty\romand\,x^2\,e^{-x^2}x^{1+2(\delta_{\romani,2}+\delta_{\romanj,2})}
  D_{\romani\romanj}(r,\nu^*,\Omega)\,,
\label{EQ_CONVOLUTION}
\end{equation}
where $x\!=\!v/v_{\romant\romanh}^\romanb$ is the particle velocity
normalized by the thermal velocity of species $\romanb\,$;
$\delta_{\romani,2}$ is the Kronecker delta. The monoenergetic
coefficients $D_{\romani\romanj}$ depend on the radially local
magnetic field strength, and are calculated and stored in a database
for fixed values of the collisionality $\nu^*\!=\!\nu R/v\iota$ and
the {\em electric field parameter}
$\Omega\!\equiv\!E_\romanr/(vB_0)$. Here, $\nu$ is the collision
frequency (see Ref.\cite{beidler2011ICNTS} for its explicit form), $R$
the major radius of the device and $\iota$ the rotational
transform. There is one independent simulation for each value of $r$,
$\nu^*$ and $\Omega$, from which we obtain the three independent
monoenergetic transport coefficients $D_{11}$, $D_{31}$ and $D_{33}$ (note that
$D_{11}\!=\!D_{12}\!=\!D_{21}\!=\!D_{22}$ and
$D_{31}\!=\!D_{32}\!=\!D_{13}\!=\!D_{23}$).\\

Following Eqs.~(\ref{EQ_LINEAR2}) and (\ref{EQ_CONVOLUTION}), one
needs to compute and convolute the monoenergetic bootstrap coefficient
$D_{31}$ in order to calculate the thermal coefficients
$L^\romanb_{31}$ and $L^\romanb_{32}$ which relate the parallel
current to the thermodynamic forces. This convolution must be done for
a given radial electric field $E_\romanr$, that is obtained from the
ambipolar condition
$\langle\Gamma_\romane\rangle(E_\romanr)\!=\!\langle\Gamma_\romani\rangle
(E_\romanr)$. Thus, according to Eq.~(\ref{EQ_LINEAR1}), the radial
diffusion coefficient $D_{11}$ must also be computed in order to
estimate $L^\romanb_{11}$ and $L^\romanb_{12}$, which relate the
radial particle fluxes to the thermodynamic forces.

Finally, the Lorentz collision operator does not conserve
momentum. For this reason, momentum-correction techniques are included
in the calculation, following method 2 in
Ref.~\cite{maassberg2009momentum}. Then, the linear relation of
Eqs.~(\ref{EQ_LINEAR1}) and ~(\ref{EQ_LINEAR2}) is not correct anymore
(but still allows the discussion of the qualitative behaviour of the
bootstrap current in terms of the transport coefficients and the
plasma gradients). For these techniques, one needs also to calculate
the parallel conductivity coefficient $D_{33}$.\\

One must note that the local ansatz underlying this approach is only
partially fulfilled at certain positions of
TJ-II~\cite{tribaldos2005global,velasco2009finite}. This makes the diffusive picture
of transport only approximately valid. Nevertheless, relaxation of the
local ansatz would make the calculation of the bootstrap current
profile impossible in terms of computing time. Also the monoenergetic
picture~\cite{lotz1988montecarlo} breaks down for very large electric
fields.

\subsection{The $\delta f$ method}\label{SEC_METHOD}

Solution of Eq.~(\ref{EQ_DKE}) in order to calculate the bootstrap
current is a numerical challenge. The reason is that one has to
calculate an asymmetry in the particle distribution function caused by
a combination of plasma radial gradients and particle trapping between
local maxima of the magnetic field strength. Although the trapped
particles do not provide a large current, the passing particle
population equilibrate with them via collisions. Then, when it comes
the time to calculate this effect by estimating $D_\mathrm{ij}$ by
means of a Monte-Carlo code, one has to be able to estimate the
non-compensation of the current carried by co-passing particles and
that carried by counter-passing particles. Note that the
trapped-particle fraction is a radially local quantity determined by
the variation of the magnetic field strength in the local flux-surface
labelled by $r$. Therefore, the dependence of the transport
coefficients (and specifically of the bootstrap coefficient) on the
trapped-particle fraction is implicitly accounted for in
$D_\mathrm{ij}(r,\nu^*,\Omega)$ of Eq.(\ref{EQ_CONVOLUTION}).

Additionally, in stellarators, particle trapping may be toroidal or
helical~\cite{peeters2000bootstrap}, and these two components may add
one to the other or cancel depending on the magnetic configuration.

Finally, in $\delta f$ Monte Carlo methods, the contribution of each
test particle to the monoenergetic coefficient is proportional to its
radial excursion from the original magnetic surface. Therefore, the
contribution of the trapped particles consists of relatively large
terms which numerically cancel out. This is the issue that is
addressed with NEO-MC. 

NEO-MC~\cite{allmaier2008variance} is an improved $\delta f$ Monte Carlo method
developed for computation of mono-energetic neoclassical transport
coefficients in stellarators. This method uses a re-discretization
procedure and importance sampling in order to reduce the variance of
these coefficients. As a result, the CPU time required for a given
accuracy of the computation of the bootstrap coefficient scales as the
mean free path to the power of 3/2. This method introduces no bias
from filtering out particles. In addition, it allows simultaneous
computation of the bootstrap coefficient and the diffusion
coefficient. NEO-MC has been benchmarked with other Monte Carlo and
field-line-tracing methods and with DKES, for different magnetic
configurations and for a range of values of the collisionality and the
electric field~\cite{beidler2011ICNTS}. The results are in good agreement,
and we make use of both NEO-MC and DKES in the following.  \\

\subsection{Error estimate}\label{SEC_ERROR}

Although NEO-MC (and DKES for not too low collisionalities) provide
small error bars, the total current may be close to zero: since the
helical and the toroidal contributions, as in every classical
stellarator, have opposite signs, they may exactly cancel out at a
given collisionality. Therefore some error estimate is required in
order to assess if the uncertainty in $D_{\romani\romanj}$ renders the
calculated currents inaccurate. In this section, we show a method for
doing so.

Although Eq.~(\ref{EQ_DKE}) is linear, $v_\romang$ includes the radial
electric field, which we calculate by means of the ambipolar condition
by root-finding (we select among multiple roots according to a
thermodynamic condition, see e.g.~\cite{turkin2011predictive}). Therefore
$L_{\romani\romanj} (E_\romanr)$ are employed to calculate $E_\romanr$
itself. In this non-linear calculation, error propagation from the
monoenergetic coefficients to the bootstrap current profile may become
cumbersome.

 We therefore undertake a Monte Carlo error propagation. We start from
a database of monoenergetic transport coefficients with their
corresponding error bars, which we have previously calculated with
NEO-MC or DKES. We then take a {\em sample} of the database: at every
radial position $\rho$, for every value of $\nu^*$, and $\Omega$, we
give numerical values to $D_{11}$, $D_{31}$ and $D_{33}$. We do
so by generating, at every point in the
($\rho$,$\nu^*$,$\Omega$)-space, three independent Gaussian random
numbers according to the average values and error bars of the
coefficients stored at the database. Then, convolution,
momentum-conserving and root-finding techniques provide the ambipolar
electric field and the neoclassical fluxes for this particular {\em
realization} of the neoclassical database. Averaging over 50 of such
samples is usually enough to present the bootstrap current profile
with a meaningful variance.  Here,
$\rho\!=\!\sqrt{\psi/\psi_0}\!\equiv\!r/a$ is the normalized radial
coordinate, where $\psi_0$ is the toroidal flux through the last
closed magnetic surface.

\begin{figure}
\begin{center}
\includegraphics[angle=270,width=0.95\columnwidth]{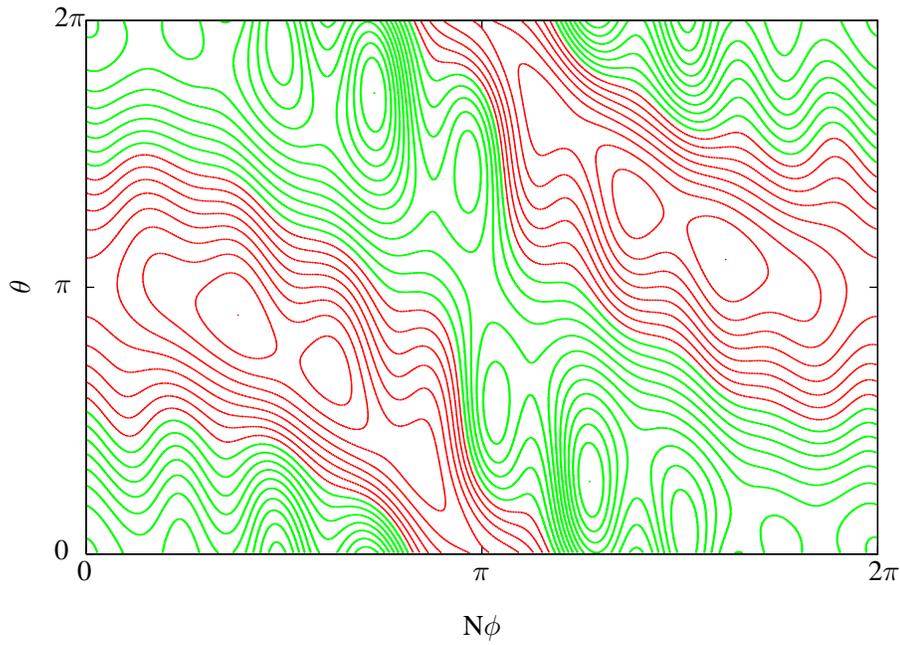}
\end{center}
\caption{Contour plot of $B$ at $\rho\!=\!0.5$. Red (black) corresponds
to $B/B_0\!>\!1$ and green (grey) corresponds to $B/B_0\!<\!1$. Two
consecutive lines differ by 0.015.}
\label{FIG_BLEVEL}
\end{figure}

\begin{figure}
\begin{center}
\includegraphics[angle=270,width=0.95\columnwidth]{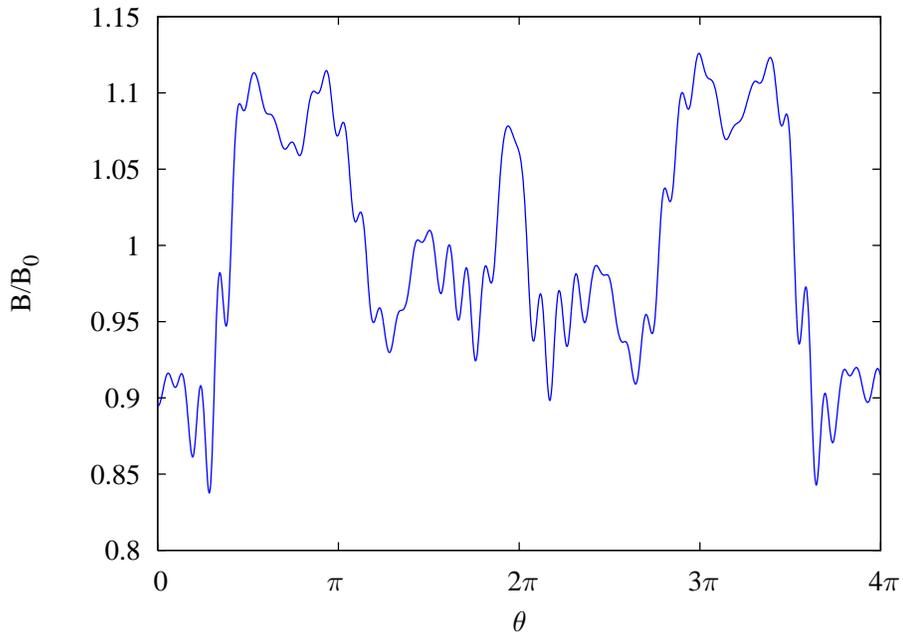}
\end{center}
\caption{Variation of $B/B_0$ along the magnetic field line passing through $\theta\!=\!0\,$
$\phi\!=\!0$ at $\rho\!=\!0.5\,$.}
\label{FIG_BLINE}
\end{figure}

 This procedure may also be straightforwardly extended to account for
the uncertainty in plasma profiles, although, as we show below, we do
not consider error bars in our density and temperatures in this work.
\\

\section{Calculation and results}\label{SEC_RESULTS}

\subsection{Magnetic configuration}\label{SEC_CONF}

TJ-II is a medium size $N\!=\!4$ heliac with bean-shaped plasma
cross-section and strong helical variation of its magnetic axis. The
magnetic configuration employed for these calculations is the
so-called 100\_44\_64, the most often employed during the experimental
campaign of TJ-II. The major radius of this configuration is
$R\!=\!1.504\,$m, the minor radius is $a\!=\!0.192\,$m and the
volume-averaged magnetic field is $0.957\,$T. The iota profile is
fairly flat, with $\iota(0)\!=\!-1.551$ and $\iota(a)\!=\!-1.650$. The
vacuum equilibrium is used here. In Fig.~\ref{FIG_BLEVEL} we plot the
contours of constant magnetic field strength $B$ at the flux-surface
given by $\rho\!=\!0.5$, where $\theta$ and $\phi$ are the poloidal
and toroidal angles in Boozer coordinates. The structure of deep
localized minima and maxima is apparent. Fig.~\ref{FIG_BLINE} show the
variation of $B$ along a field line starting at $\theta\!=\!0\,$
$\phi\!=\!0$, which clearly cannot be described with a few Fourier
modes. Hence the convergence problems of DKES.

\begin{figure}
\begin{center}
\includegraphics[angle=270,width=1\columnwidth]{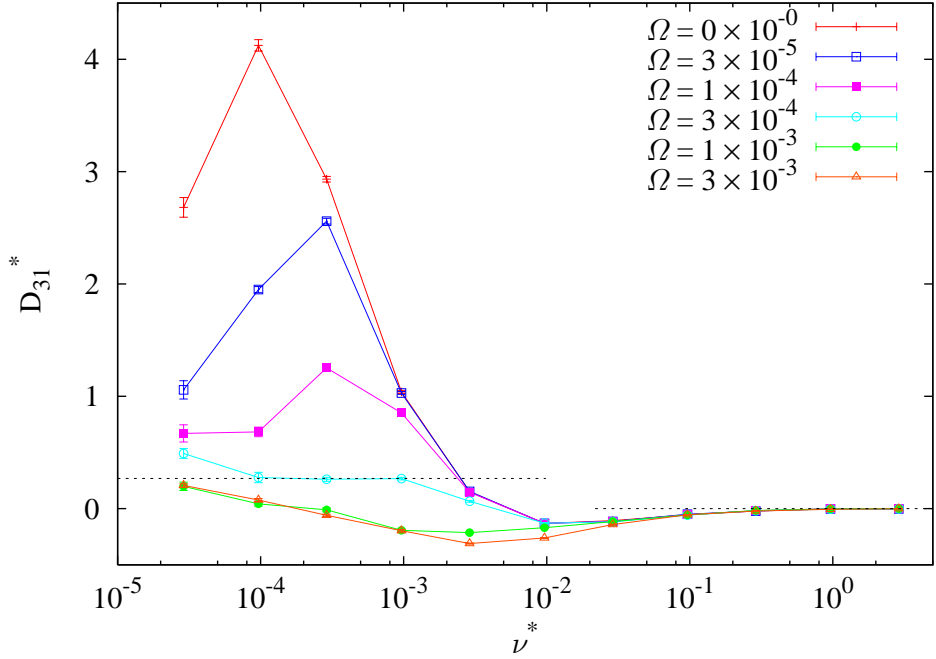}
\end{center}
\caption{Monoenergetic bootstrap coefficient at $\rho\!=\!0.5$ as a
function of the collisionality for different electric fields. The
dashed lines represent the $\nu^*\to0$ and $\nu^*\to\infty$ limits.}
\label{FIG_MONO}
\end{figure}

\subsection{Monoenergetic coefficients}\label{SEC_MONO}

We have calculated, for the configuration described in the previous
subsection, the monoenergetic coefficient $D_{31}$, together with
$D_{11}$ and $D_{33}$ for a wide range of the collisionality and the
electric field parameter. We calculate the dependence of the
monoenergetic coefficient with $\nu^*$ (between $3\times 10^{-5}$ and
$3\times 10 ^2$) for fixed values of $\Omega$ between 0 and $1\times
10^{-1}$. We have repeated these calculations at several radial
locations, from $\rho\!=\!0.08$ to $\rho\!=\!0.98$.

Part of the results, at middle radial position ($\rho\!=\!0.5$) are
shown in Fig.~\ref{FIG_MONO}. The monoenergetic coefficient is
presented there normalized to that of an axisymmetric tokamak of the
same aspect ratio in the banana regime:
\begin{equation}
  D^*_{31}\equiv D_{31}/D^\mathrm{b}_{31}\,,\quad
  D^\mathrm{b}_{31}=1.46\frac{m_\romanb v^2}{3B_0\iota\sqrt{\varepsilon}Z_\romanb e}\,,
\end{equation}
where $\varepsilon$ is the inverse aspect ratio and $m_\romanb$ the
particle mass. The results for $D^*_{31}$ are qualitatively similar to
those observed for LHD, CHS, W7-AS and others~\cite{beidler2011ICNTS}:
for $\nu^*\!>\!1$, the collision frequency $\nu$ is larger than the
bounce frequency, hence there are no trapped particles and no seeding
mechanism for the bootstrap current. Note that the monoenergetic
coefficient is already close to zero for lower collisionalities than
$\nu^*\!=\!1$, for the bounce frequency is smaller than the {\em
effective-detrapping collision frequency}, which may be defined as
$\nu/\varepsilon$. For very low collisionalities, as the ratio between
the electric field $\Omega$ and the collisionality $\nu^*$ increases,
the coefficient converges to an asymptotic value depending on the
configuration and the radial
position~\cite{shaing1983neoclassical,nakajima1989bootstrap}. The
predicted asymptotic limit has been calculated by means of field-line
integration~\cite{kernbichler2002neo}.

Low error bars are obtained in the whole range of collisionalities
calculated while, for this configuration, DKES cannot provide accurate
values for collisionalities lower than $3\times 10^{-4}$. Actually,
these small error bars push the convergence with collisionless
asymptote towards collisionalities even lower than $3\times
10^{-5}$. In the case of very small (especially zero) electric field
parameter, this seems to happen at collisionalities that are
completely out of reach for our computing resources. Nevertheless, for
the collisionalities of the plasmas presented in
Section~\ref{SEC_PROFILE}, the data of Fig.~\ref{FIG_MONO} allow for a
precise calculation of the bootstrap current. Only under reactor
conditions ($n\sim3\times 10^{20} m^{-3}$, $T\sim20\,$keV) would
$\nu^*\sim 5\times 10^{-5}$ --- $10^{-4}$ correspond to thermal particles.

Additionally, for some values of $\Omega$ such as $3\times 10^{-4}$,
the monoenergetic coefficients do not converge to the collisionless
limit. This effect was already been observed in Ref.~\cite{beidler2011ICNTS},
and is still to be understood. Nevertheless, one has to keep in mind
that for large values of the electric field, the starting hypothesis
of the calculation are not fulfilled
anymore~\cite{beidler2007icnts,velasco2009finite}: the kinetic energy is not an
approximately conserved quantity. Fortunately, the electric field in
this region of TJ-II is low enough, specially for NBI plasmas, and
these data will not be required.
\\

Special attention must be paid to the interpolation and extrapolation of
the monoenergetic database and to the calculation of the integral in
Eq.~(\ref{EQ_CONVOLUTION}). This integration is made by means of
Gauss-Laguerre of order 64, and interpolation by 3-point Lagrange,
with $\nu^*$ and $\Omega$ in logarithmic scale. In extrapolation, as
in Ref.~\cite{spong2005flow}, we have made use of the asymptotic limit
where available, but also tried extrapolations such as
$D^*_{31}(\nu^*<3\times 10^{-5})\!=\!D^*_{31}(\nu^* =3\times 10^{-5})$. We
have made sure that none of these choices affects the final result.

Once we have fully covered the relevant range of collisionalities with
enough accuracy, the next step in the calculation of the bootstrap
coefficient is to convolute the results shown in
Section~\ref{SEC_MONO} with a Maxwellian distribution function. Then,
the solution of a linear system of moment equations for which the
coefficients are different energy moments of the mono-energetic
transport coefficients~\cite{maassberg2009momentum} allows for the
calculation of the ambipolar electric field and then the bootstrap
current.  \\

\subsection{Current profile}\label{SEC_PROFILE}

The bootstrap current has been estimated to be of the order of $1\,$kA
for low density plasmas of
TJ-II~\cite{tribaldos2003bootstrap,estrada2002transient}. Since the
magnetic field of TJ-II has a very broad Fourier spectrum, calculation
in the {\em lmfp} regime is very complicated, and the results
presented large error bars for low
collisionalities~\cite{ribeiro1987plateau,yunta1990bootstrap,tribaldos2003bootstrap}. It
has been argued that, due to the exponential in
Eq.~(\ref{EQ_CONVOLUTION}), the unacceptably large error bars for
$\nu^*\sim 10^{-4}$, well into the {\em lmfp} might not preclude the
calculation of the bootstrap current~\cite{tribaldos2003bootstrap},
but this was not quantified as we propose in
Section~\ref{SEC_ERROR}. These neoclassical calculations were compared
to the measurements of the toroidal rotation velocity of some protons
and impurities by means of passive emission
spectroscopy~\cite{rapisarda2005rotation}. Although there was some
qualitative agreement, the results differed by a factor 2.  \\

Recently, lithium wall coating at TJ-II~\cite{sanchez2009transitions} has given
access to low impurity concentration and high density at plasmas
heated with Neutral Beam Injection. For these plasmas, in which the
{\em lmfp} regime is less relevant, the bootstrap current should be
easier to calculate. Nevertheless, being smaller in magnitude, the
significance of the results must be checked.

\begin{figure}
\begin{center}
\includegraphics[angle=270,width=0.99\columnwidth]{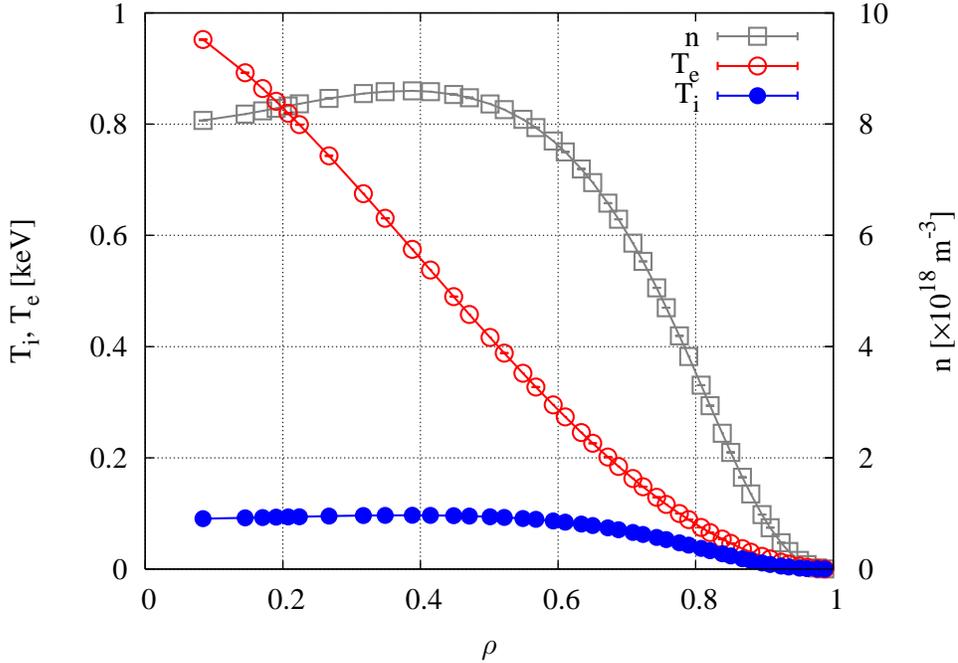}
\end{center}
\caption{Plasma profiles for a low-density plasma of TJ-II.}
\label{FIG_LD}
\end{figure}

\begin{figure}
\begin{center}
\includegraphics[angle=270,width=0.99\columnwidth]{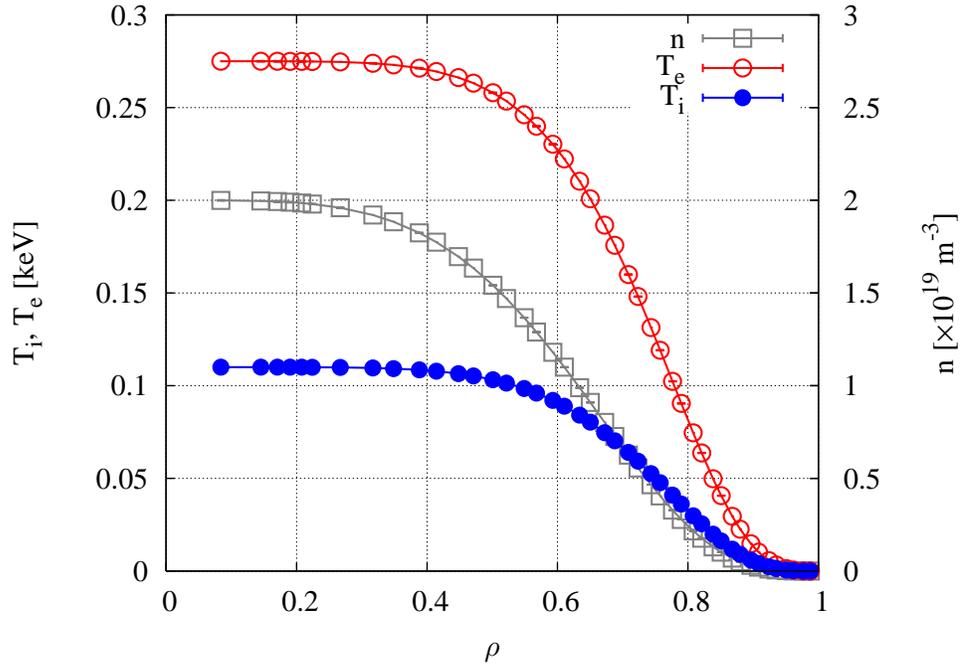}
\end{center}
\caption{Plasma profiles for a medium-density plasma of TJ-II.}
\label{FIG_MD}
\end{figure}

\begin{figure}
\begin{center}
\includegraphics[angle=270,width=0.99\columnwidth]{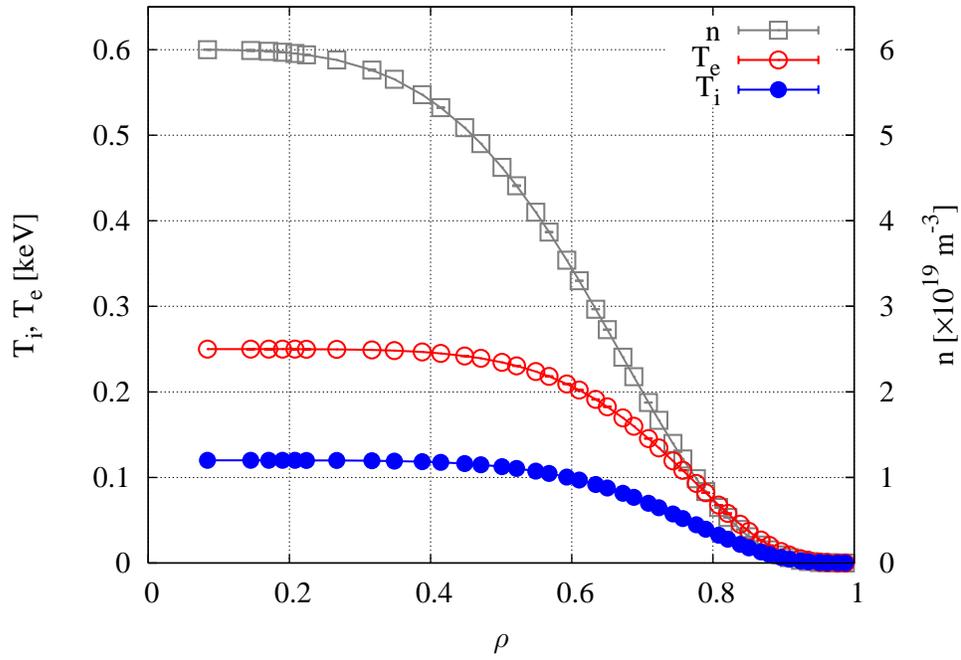}
\end{center}
\caption{Plasma profiles for a high-density plasma of TJ-II.}
\label{FIG_HD}
\end{figure}

We calculate the bootstrap current profile for three regimes of TJ-II,
corresponding to low, medium and high density.
Figs.~\ref{FIG_LD},~\ref{FIG_MD} and~\ref{FIG_HD} show plasma profiles
characteristic of the three regimes. Low density plasmas at TJ-II have
peaked electron temperature profiles, with central values of above
1$\,$keV and hollow density profiles. As the density rises, both
profiles become flatter and the electron temperature becomes lower, of
about 200 ---300$\,$eV for high density plasmas heated by means of
Neutral Beam Injection. The ion temperature is rather flat under all
conditions, of the order of 100$\,$eV. For these plasmas with
lithium-coated walls, taking the effective charge $Z_{eff}\!=\!1\,$ is
a reasonable approximation. Note that, since for thermal particles
$\nu^*\propto n/T^2\,$, higher density means higher collisionality,
for the temperature decreases (for electrons) or barely changes
(for ions).  \\

\begin{figure}
\begin{center}
\includegraphics[angle=270,width=0.9\columnwidth]{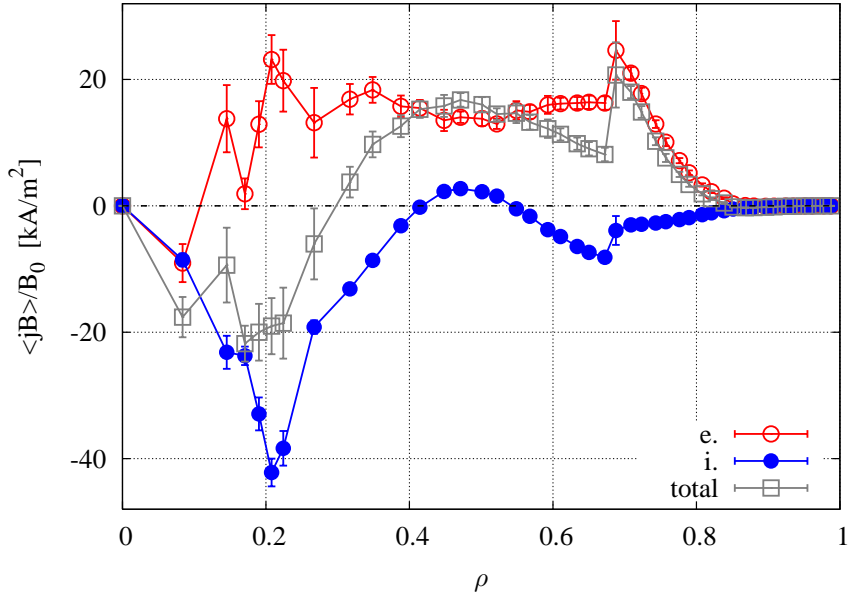}
\end{center}
\caption{Calculated bootstrap current profile for the low density
 plasma. The error bars are calculated following
 Section~\ref{SEC_ERROR}}
\label{FIG_JLD}
\end{figure}

\begin{figure}
\begin{center}
\includegraphics[angle=270,width=0.9\columnwidth]{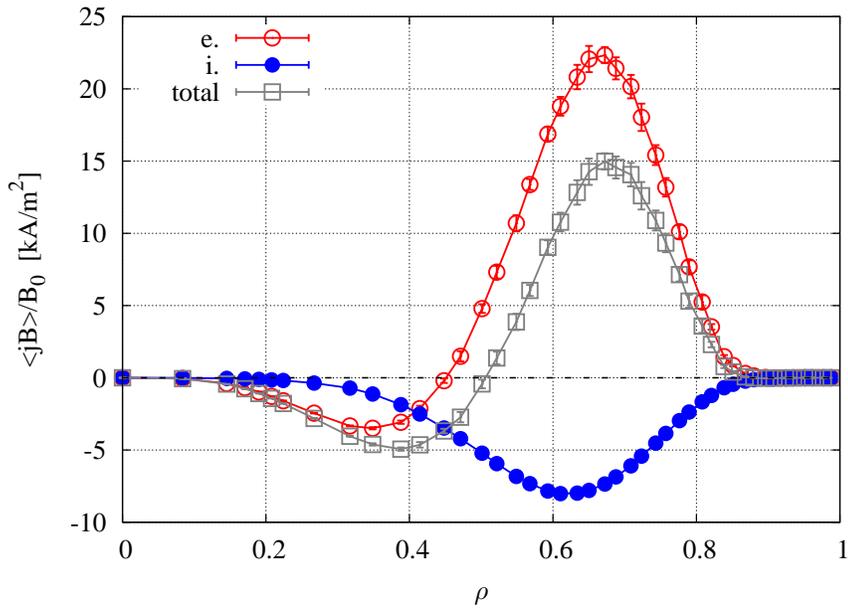}
\end{center}
\caption{Calculated bootstrap current profile for the medium density plasma.}
\label{FIG_JMD}
\end{figure}

\begin{figure}
\begin{center}
\includegraphics[angle=270,width=0.9\columnwidth]{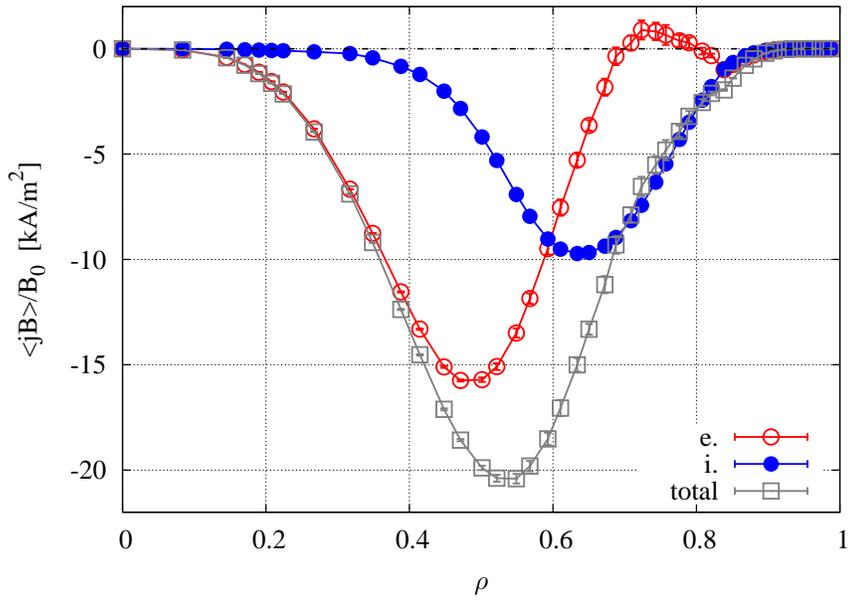}
\end{center}
\caption{Calculated bootstrap current profile for the high density plasma.}
\label{FIG_JHD}
\end{figure}

We show the results in Figs.~\ref{FIG_JLD},~\ref{FIG_JMD}
and~\ref{FIG_JHD}, separating the contributions from electrons and
ions. The latter may be compared to toroidal rotation measurements at
TJ-II~\cite{rapisarda2005rotation}, available for low density plasmas.

Some general features may be extracted from observation of the three
figures. First of all, the current is mainly carried by the
electrons. This is a consequence of the decoupling of the ion and
electron temperatures, the latter being systematically higher, in
TJ-II plasmas. Thus, the electrons are much less collisional and their
contribution is much higher. Also, for the high-density cases (and the
low density at outer positions), the plasma is in the {\em ion
root}~\cite{mynick1983roots}: a negative radial electric field
partially cancels out the ion channel while adding an additional
contribution to the electron channel, see Eqs.~(\ref{EQ_LINEAR1}) and
(\ref{EQ_LINEAR2}). The exception are the inner radial positions in
the low-density case, where the electron root is realized. There, the
positive radial electric field drives a relatively high ion parallel
flow (with low damping from the small trapped-particle friction) and
the electron flow is coupled to it~\cite{maassberg2009momentum}. Also
near the gradient zone of high density plasmas the temperatures are
similar and the ion bootstrap current is not negligible.

The maxima of the bootstrap current density follow those of the
temperature gradient, since $L^\romane_{32}-1.5\,L^\romane_{31}$ is
systematically much higher than $L^\romane_{31}$. This reflects the
{\em overshoot} in the bootstrap coefficient for collisionalities
between $10^{-4}$ and $10^{-3}$, a usual feature in
stellarators~\cite{beidler2011ICNTS}. The higher the electron collisionality,
the lower the amplitude of the maxima.

There is a switch in the sign of the bootstrap current at medium
radius, corresponding to a switch in the sign of the electron
contribution. It may be traced-back as the change of sign in the
bootstrap coefficient, corresponding to the cancellation of the
stellarator-like and the tokamak-like contributions. This happens for
a given collisionality at each radial position (at medium radius,
around $\nu^*\!=\!=3\times 10^{-3}$, see Fig.\ref{FIG_MONO}). For
higher densities, the switch in sign happens for larger minor
radius. The consequence of this fact is that the total toroidal
current is positive (tokamak-like) for low density plasmas and becomes
negative (stellarator-like) for high densities. Integration of
Figs.~\ref{FIG_JLD},~\ref{FIG_JMD} and~\ref{FIG_JHD} yield the
following total currents, with the error bars in parentheses:
$0.55(2)\,$kA for the ECH plasma, $0.314(8)\,$kA for the medium
density NBI plasma and $-0.739(3)\,$kA for the high density
plasma. This change of sign in the toroidal current has been observed
in TJ-II operation, and the values obtained here are comparable to
that measured by Rogowski coils~\cite{estrada2002transient}. 

The current in Fig.~\ref{FIG_JLD} presents a discontinuous behaviour
around $\rho\!=\!0.69$. It corresponds to the transition from electron
to ion root: two stable roots exist from $\rho\!=\!0.63$ to
$\rho\!=\!0.71$ and the electric field is evaluated by root-finding,
so the transition layer is very narrow. If a differential equation for
the electric field~\cite{turkin2011predictive} were solved instead,
the transition layer would be broader and Fig.~\ref{FIG_JLD} would
become smoother. The qualitative results would be the same, with a
total toroidal current ranging between $0.5$ and $0.75\,$kA. These
results remain also of significance if error bars of 20\% are allowed
in the density and temperature profiles.

A consequence of the high collisionality of the ions is that their
contribution to the bootstrap current suffers from very low error
bars, while those of the electron contribution are much
higher. Therefore, comparison of these results with measurements by
Charge-Exchange recombination spectroscopy (CXRS) is to be
done. Actually, much of the error bars in
Figs.~\ref{FIG_JLD},~\ref{FIG_JMD},~\ref{FIG_JHD} comes from the
uncertainty in $D_{33}$ coefficient for low collisionality, which is
calculated with DKES and employed in the momentum-correction
techniques. Although still the results are of significance, the
calculation of $D_{33}$ with NEO-MC is underway.  \\

\begin{figure}
\begin{center}
\includegraphics[angle=270,width=1\columnwidth]{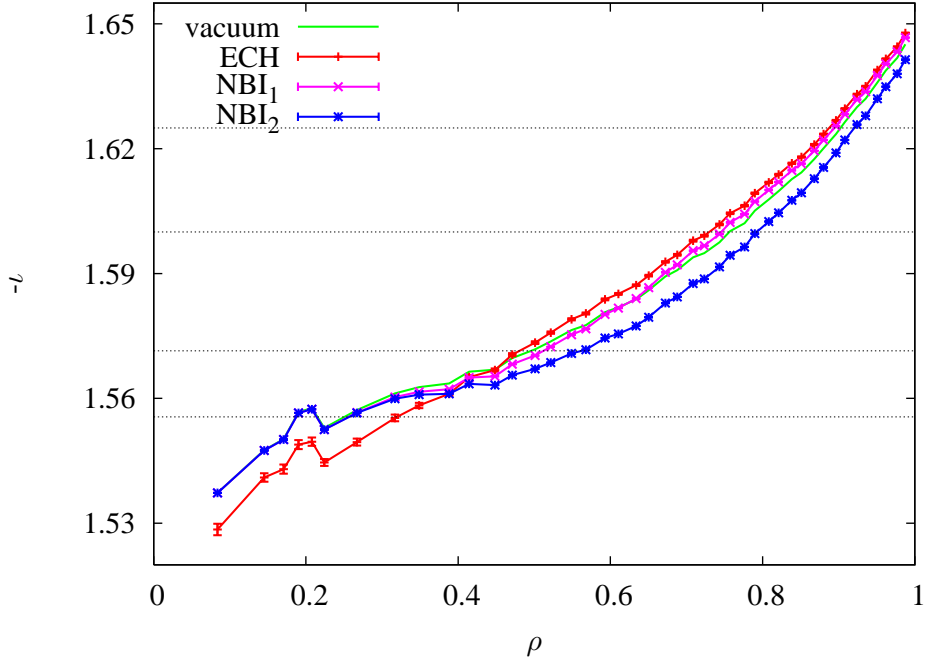}
\end{center}
\caption{Rotational transform profile corresponding to the vacuum
configuration and the three plasma profiles.}
\label{FIG_IOTA}
\end{figure}

In Fig.~\ref{FIG_IOTA} we show an estimate of how the bootstrap
current modifies the rotational transform profile. We also plot the
rationals present in the plasma for the 100\_44\_64 configuration. In
the vacuum iota range, calculated with VMEC~\cite{hirshman1986vmec},
several rationals (14/9, 11/7, 8/5 and 13/8) are present. Since the
total toroidal current is small in absolute value for TJ-II, the most
external rational barely moves. Inner rationals do move for ECH
plasmas, which have plasma gradients for small minor radius. Note that
negative currents lead to reduced $|\iota|$, and hence unwind the
rotational transform.\\

The calculation presented is a linear approximation to a non-linear
problem: this current is, as stated before, the result of the
combination of the non-uniformity of the magnetic field in the flux
surfaces and the density and temperature profiles. Since the
calculated current is not zero, it will modify the vacuum magnetic
field. The calculation of the bootstrap current should be iterated in
the new magnetic configuration, and the process should be repeated
until one achieves convergence (i.e., the calculated current does not
vary from one iteration to the next). This type of calculation has
been made for LHD~\cite{watanabe1992lhd}, which is a high-$\beta$
device with relatively large bootstrap current and large Shafranov
shift. It has also been performed for simplified models of NCSX and
QHS~\cite{ferrando2004qas}, in order to study the preservation of the
quasi-symmetry. Since TJ-II lacks any of these quasi-symmetries, is
designed to have small Shafranov shift, and the pressures and
calculated currents are low, it is not necessary to iterate further.
\\

\begin{table}
\begin{center}
\begin{tabular}{ccccc}
  Configuration name  & $\langle B(T)\rangle_{vol}$ & V($m^3$) & $\iota(0)$ & $\iota(a)$
  \\\hline\hline
100\_32\_60 & 1.087 & 0.934 & -1.423 & -1.517 \\
100\_38\_62 & 0.971 & 1.031 & -1.492 & -1.593 \\
100\_40\_63 & 0.960 & 1.043 & -1.510 & -1.609 \\
100\_42\_63 & 0.931 & 1.079 & -1.534 & -1.630 \\
100\_44\_64 & 0.962 & 1.098 & -1.551 & -1.650 \\
100\_46\_65 & 0.903 & 1.092 & -1.575 & -1.676 \\
100\_50\_65 & 0.962 & 1.082 & -1.614 & -1.704 \\
100\_55\_67 & 0.964 & 1.073 & -1.659 & -1.739 \\
\hline
\end{tabular}
\end{center}
\caption{Main parameters of the configuration scan}
\label{TAB_CONF}
\end{table}

\begin{figure}
\begin{center}
\includegraphics[angle=270,width=0.98\columnwidth]{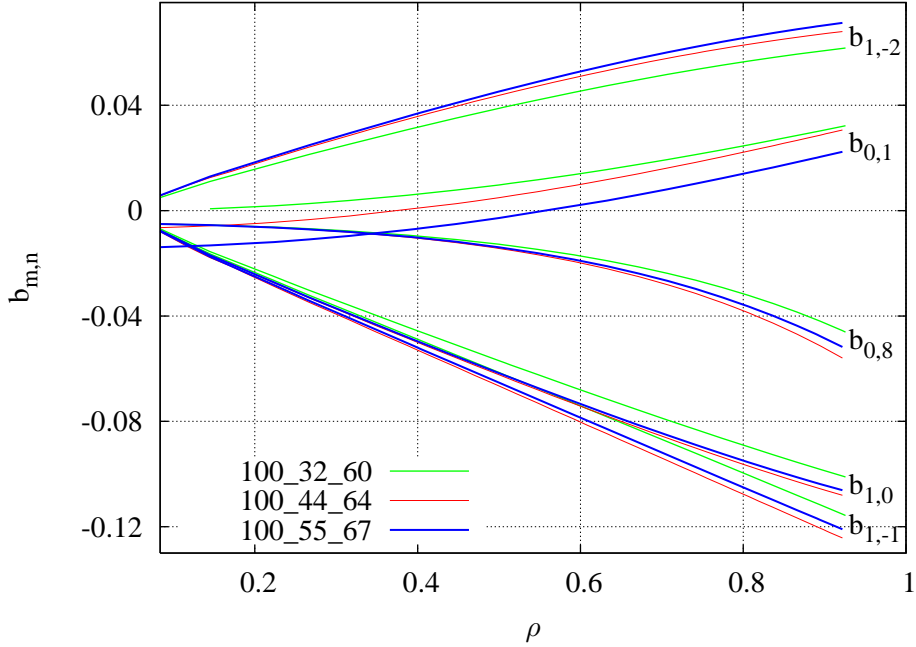}
\end{center}
\caption{Main Fourier harmonics of the configurations 100\_32\_60, 100\_44\_64, 100\_55\_67}
\label{FIG_FOURIER}
\end{figure}

\begin{figure}
\begin{center}
\includegraphics[angle=270,width=1\columnwidth]{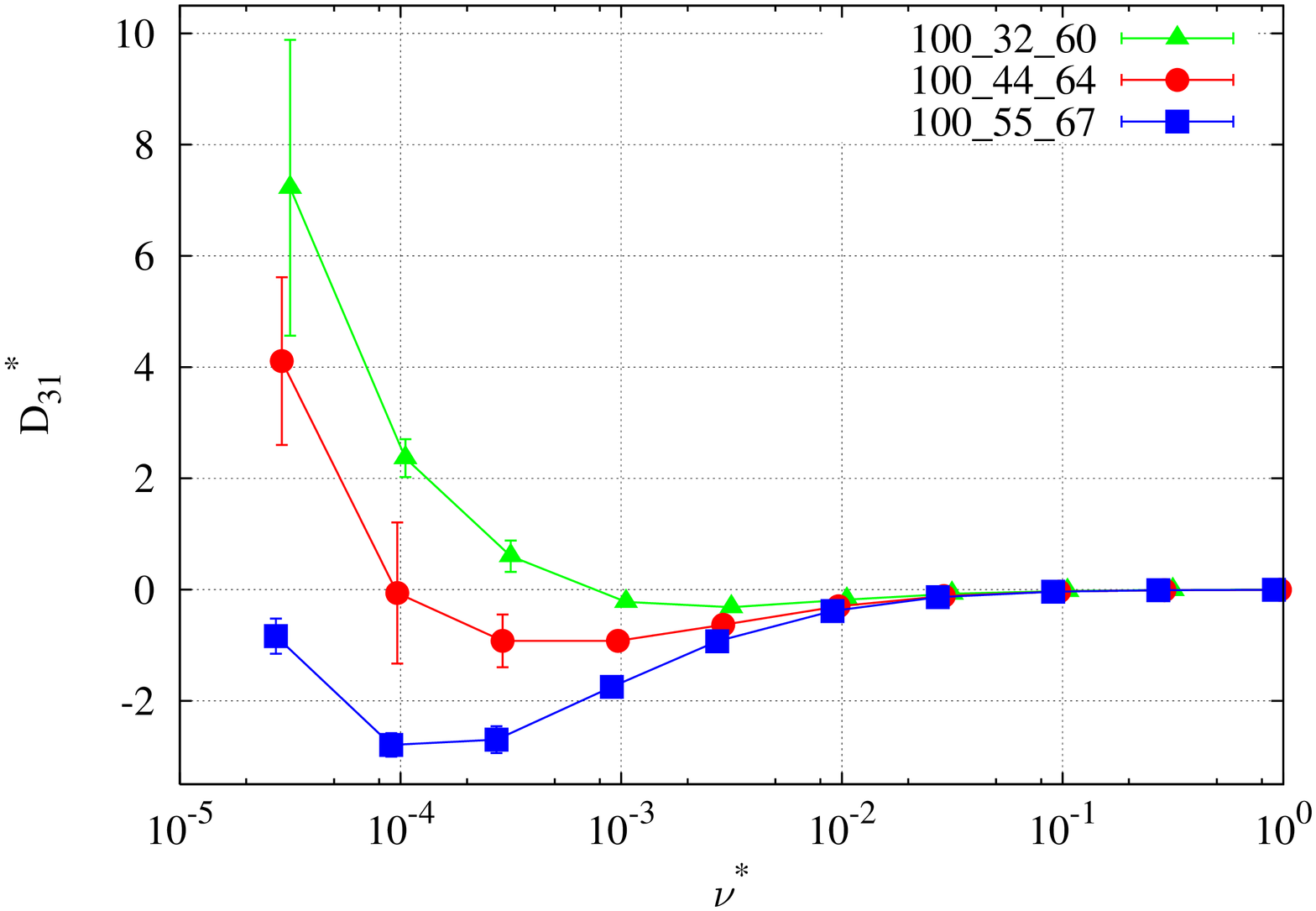}
\end{center}
\caption{Bootstrap current coefficient at $\rho\!=\!0.25$ as a
function of the collisionality}.
\label{FIG_CONF025}
\end{figure}

\begin{figure}
\begin{center}
\includegraphics[angle=270,width=1\columnwidth]{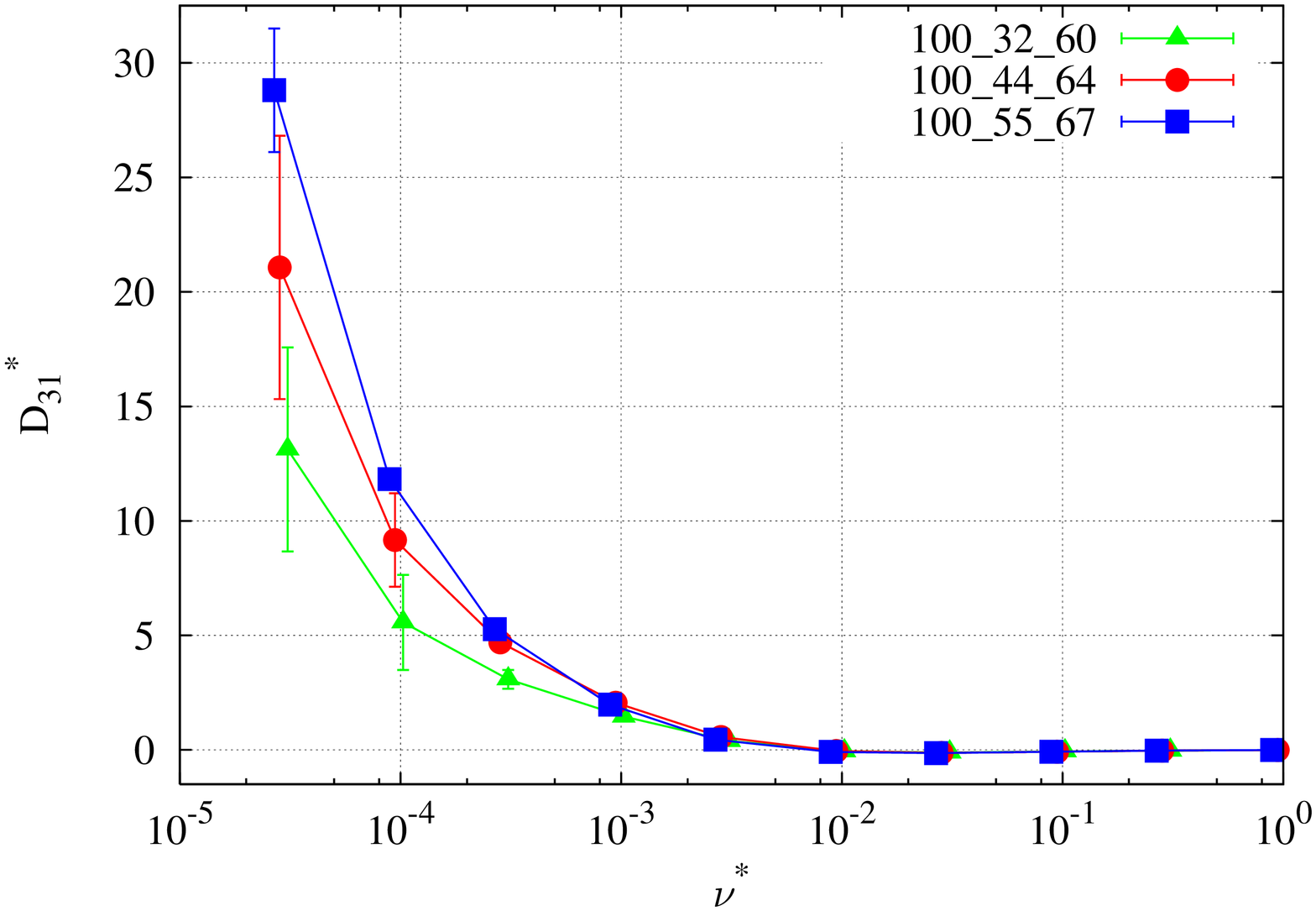}
\end{center}
\caption{Bootstrap current coefficient at $\rho\!=\!0.7$ as a function of
the collisionality.}
\label{FIG_CONF070}
\end{figure}

\subsection{Configuration dependence of the bootstrap current}\label{SEC_CONFS}

Finally, we make a scan in a relevant part the configuration space of
TJ-II. A thorough study of the dependence of the parallel fluxes on
the magnetic configuration is beyond the scope of this work; and also
NEO-MC computations, although feasible, are quite time consuming. We
thus try to describe the bootstrap current of the configuration in
terms of its monoenergetic coefficient at $\rho\!=\!0.25$ and
$\rho\!=\!0.7$ (where the electron temperature gradients are larger
for low and high density plasmas respectively) for $\Omega\!=\!0$. We
calculate so for seven different configurations which are regularly
explored in TJ-II operation. We show in Table.~\ref{TAB_CONF} the main
global parameters: when switching from one configuration to another,
the volume is held approximately constant, while the rotational
transform profile is raised without changing the shear. The
neoclassical transport of these configurations is not expected to
change much, since the main Fourier terms in the description of the
magnetic field remain unchanged: Fig.~\ref{FIG_FOURIER} shows the
radial profile of the main terms in the Fourier decomposition in
Boozer coordinates(see Ref.~\cite{solano1988tj-ii} for a more
comprehensive discussion), given by
\begin{equation}
B(\rho,\phi,\theta)/B_0(\rho)=\sum_{n=-\infty}^{\infty}\sum_{m=0}^{\infty}b_{m,n}(\rho)cos(m\theta-Nn\phi)\,.
\label{EQ_FOURIER}
\end{equation}
For the sake of clarity, only the standard 100\_44\_64 configuration
and the extremes of the scan (100\_32\_60 and 100\_55\_67), are
presented in the plot. The two main Fourier components relevant for
the bootstrap coefficient are the helical curvature $b_{1,-1}$ and the
toroidal curvature $b_{1,0}$. These components vary a few percents
during the configuration scan. The next contribution may come from the
toroidal mirror $b_{0,1}$~\cite{maassberg2005ecd,nishimura2009viscosity}. This is the
component of largest relative variation during the magnetic scan, even
reversing its sign. Nevertheless, $b_{0,1}$ is small compared with the
two curvatures and therefore it does not have a large absolute impact
on the bootstrap current, as we will see below. The relative variation
in the remaining components of the Fourier spectrum is negligible.

The calculation is made with DKES: up to 150 Legendre polynomials and
up to 2548 Fourier modes have been employed in the description of the
distribution function and the largest 50 Fourier modes have been kept
in the description of the magnetic field on every surface. We have
made sure that increasing the number of modes does not modify the
presented results within the error bars.

In Figs.~\ref{FIG_CONF025} and~\ref{FIG_CONF070} we show the
collisionality dependence of the bootstrap coefficient for three
different configurations at $\rho\!=\!0.25$ and $\rho\!=\!0.7$. For
the sake of clarity, only configurations 100\_32\_60, 100\_44\_64 and
100\_55\_67 will be shown in the figures. Since the $D^*_{31}$
coefficient has already been discussed, we focus on the differences
between configurations. Although the calculations suffer from large
error bars for low collisionalities, one may see that the normalized
coefficients coincide for large collisionality and differ only for
small collisionalities. The differences are relatively (but not
absolutely) larger for small minor radius. At this region, the
toroidal and helical curvatures are relatively smaller, and therefore
differences in the toroidal mirror trapping show up in the bootstrap
coefficient.  \\

A $1/\iota$ dependence of the bootstrap current is thus to be expected
(since $D_{31}\!\propto\!D_{31}^*/\iota$) for the medium and
high-density plasmas of Figs.~\ref{FIG_MD} and~\ref{FIG_HD}.

For low-density plasmas, the situation is different for three reasons:
first of all, the electron collisionality is significantly lower. Also
the electron temperature gradient is large, see Fig.~\ref{FIG_LD}, so
the contribution of the term proportional to $L^\romane_{32}$ in
Eq.~(\ref{EQ_LINEAR2}) becomes larger. Finally, this gradient is large
also for inner positions, where the differences in $D^*_{31}$ are
relatively larger, see Fig.~\ref{FIG_CONF025}. The three facts tend to
maximize the difference between configurations.

Finally, the sidebands not shown in Fig.~\ref{FIG_FOURIER} might yield
different radial diffusion coefficients for different
configurations. Although this situation is not likely to happen, the
ambipolar equation would yield different radial electric fields for
given plasma profiles, and thus different bootstrap current
profiles. This calculation is left for future works.

\section{Conclusions}\label{SEC_CONCLUSIONS}

Calculations of the bootstrap current for the TJ-II stellarator are
presented. The $\delta f$ code NEO-MC, together with DKES, has allowed
for the first time for the precise computation of the neoclassical
bootstrap transport coefficient in the long mean free path regime of
the vacuum configuration of TJ-II. The low error bars have allowed for
a precise convolution of the monoenergetic coefficients. Since the
calculation has been repeated at several radial positions, we are able
for the very first time to calculate the radial profile of the
bootstrap current of several plasmas of TJ-II with sufficient accuracy
for comparison with experimental results. A Monte Carlo method for
error estimation allows us to confirm the significance of the results.
 
The current profile is well understood in terms of the plasma
profiles. Since the electrons are within the {\em lmfp} and the
plateau regime for all the plasmas considered, the gradient in the
electron temperature provides the main contribution to the total
current density.

The results show qualitative agreement with previous calculations and
  measurements at TJ-II; future work includes comparison with CXRS and
  Rogowski coils measurements for real TJ-II discharges.

We have calculated the bootstrap coefficient for several
configurations usually operated at TJ-II. The results suggest that the
dependence of the bootstrap current on the magnetic configuration
should be small, specially for NBI plasmas. A more comprehensive study
is left for a future work.

This calculation will allow to quantify more accurately finite-beta
effects in the studies on the relation between confinement and
magnetic configuration. Additionally, these results must be kept in
mind when estimating the EBW current drive at TJ-II.

\section{Acknowledgments}

The authors are grateful to S.P. Hirshman and D. Spong for the DKES
and VMEC codes, to J. Geiger for his help with the VMEC code and to
J.A. Alonso and E. Solano for useful discussions. This work was
carried out within the Associations EURATOM-CIEMAT, \"OAW-EURATOM and
IPP-EURATOM. Part of the work was funded by the Spanish Ministerio de
Ciencia e Innovaci\'on, Spain, under Project ENE2008-06082/FTN and by
the Austrian Science Foundation, FWF, under contract number
P16797-N08.

\section*{References}

\bibliographystyle{iopart-num}

\bibliography{./bibliography}

\providecommand{\newblock}{}
\begin{thebibliography}{10}
\expandafter\ifx\csname url\endcsname\relax
  \def\url#1{{\tt #1}}\fi
\expandafter\ifx\csname urlprefix\endcsname\relax\def\urlprefix{URL }\fi
\providecommand{\eprint}[2][]{\url{#2}}

\bibitem{peeters2000bootstrap}
Peeters A~G 2000 {\em Plasma Physics and Controlled Fusion\/} {\bf 42} B231
  \urlprefix\url{http://stacks.iop.org/0741-3335/42/i=12B/a=318}

\bibitem{helander2002collisional}
Helander P and Sigmar D~J 2002 {\em Collisional transport in magnetized
  plasmas\/} vol~87 (Cambridge University Press)

\bibitem{turkin2006current}
Turkin Y, Maa{\ss}berg H, Beidler C~D, Geiger J and Marushchenko N~B 2006 {\em
  Fusion Science and Technology\/} {\bf 50} 387--394
  \urlprefix\url{http://www.new.ans.org/pubs/journals/fst/a1260}

\bibitem{alejaldre1999first}
Alejaldre C, Alonso J, Almoguera L, Ascas\'ibar E, Baciero A, Balb\'in R,
  Blaumoser M, Botija J, Bra{\~n}as B, de~la Cal E, Cappa A, Carrasco R,
  Castej\'on F, Cepero J~R, Cremy C, Doncel J, Dulya C, Estrada T, Fern\'andez
  A, Franc\'es M, Fuentes C, Garc\'ia A, Garc\'ia-Cort\'es I, Guasp J, Herranz
  J, Hidalgo C, Jim\'enez J~A, Kirpitchev I, Krivenski V, Labrador I, Lapayese
  F, Likin K, Liniers M, L\'opez-Fraguas A, L\'opez-S\'anchez A, de~la Luna E,
  Mart\'in R, Mart\'inez A, Medrano M, M\'endez P, McCarthy K, Medina F, van
  Milligen B, Ochando M, Pacios L, Pastor I, Pedrosa M~A, de~la Pe{\~n}a A,
  Portas A, Qin J, Rodr\'iguez-Rodrigo L, Salas A, S\'anchez E, S\'anchez J,
  Tabar\'es F, Tafalla D, Tribaldos V, Vega J, Zurro B, Akulina D, Fedyanin
  O~I, Grebenshchicov S, Kharchev N, Meshcheryakov A, Barth R, van Dijk G,
  van~der Meiden H and Petrov S 1999 {\em Plasma Physics and Controlled
  Fusion\/} {\bf 41} A539
  \urlprefix\url{http://stacks.iop.org/0741-3335/41/i=3A/a=047}

\bibitem{alejaldre1999confinement}
Alejaldre C, Alonso J, Almoguera L, Ascas\'ibar E, Baciero A, Balb\'in R,
  Blaumoser M, Botija J, Bra{\~n}as B, de~la Cal E, Cappa A, Carrasco R,
  Castej\'on F, Cepero J~R, Cremy C, Delgado J~M, Doncel J, Dulya C, Estrada T,
  Fern\'andez A, Fuentes C, Garc\'ia A, Garc\'ia-Cort\'es I, Guasp J, Herranz
  J, Hidalgo C, Jim\'enez J~A, Kirpitchev I, Krivenski V, Labrador I, Lapayese
  F, Likin K, Liniers M, L\'opez-Fraguas A, L\'opez-S\'anchez A, de~la Luna E,
  Mart\'in R, Martinez A, Mart\'inez-Laso L, Medrano M, M\'endez P, McCarthy
  K~J, Medina F, van Milligen B, Ochando M, Pacios L, Pastor I, Pedrosa M~A,
  de~la Pe{\~n}a A, Portas A, Qin J, Rodr\'iguez-Rodrigo L, Salas A, S\'anchez
  E, S\'anchez J, Tabar\'es F, Tafalla D, Tribaldos V, Vega J, Zurro B, Akulina
  D, Fedyanin O~I, Grebenshchikov S, Kharchev N, Meshcheryakov A, Sarksian K~A,
  Barth R, van Dijk G and van~der Meiden H 1999 {\em Plasma Physics and
  Controlled Fusion\/} {\bf 41} B109
  \urlprefix\url{http://stacks.iop.org/0741-3335/41/i=12B/a=307}

\bibitem{ascasibar2002confinement}
Ascas\'ibar E, Alejaldre C, Alonso J, Almoguera L, Baciero A, Balb\'in R,
  Blanco E, Blaumoser M, Botija J, Bra{\~n}as B, Cappa A, Carrasco R,
  Castej\'on F, Cepero J~R, Chmyga A~A, Doncel J, Dreval N~B, Eguilior S,
  Eliseev L, Estrada T, Fedyanin O, Fern\'andez A, Fontdecaba J~M, Fuentes C,
  Garc\'ia A, Garc\'ia-Cort\'es I, Gonçalves B, Guasp J, Herranz J, Hidalgo A,
  Hidalgo C, Jim\'enez J~A, Kirpitchev I, Khrebtov S~M, Komarov A~D, Kozachok
  A~S, Krupnik L, Lapayese F, Likin K, Liniers M, L\'opez-Bruna D,
  L\'opez-Fraguas A, L\'opez-R\'azola J, L\'opez-S\'anchez A, de~la Luna E,
  Malaquias A, Mart\'in R, Medrano M, Melnikov A~V, M\'endez P, McCarthy K~J,
  Medina F, van Milligen B, Nedzelskiy I~S, Ochando M, Pacios L, Pastor I,
  Pedrosa M~A, de~la Pe{\~n}a A, Petrov A, Petrov S, Portas A, Romero J,
  Rodr\'iguez-Rodrigo L, Salas A, S\'anchez E, S\'anchez J, Sarksian K,
  Schchepetov S, Skvortsova N, Tabar\'es F, Tafalla D, Tribaldos V, Varandas
  C~F~A, Vega J and Zurro B 2002 {\em Plasma Physics and Controlled Fusion\/}
  {\bf 44} B307 \urlprefix\url{http://stacks.iop.org/0741-3335/44/i=12B/a=322}

\bibitem{ascasibar2005magnetic}
Ascas\'ibar E, Estrada T, Castej\'on F, L\'opez-Fraguas A, Pastor I, S\'anchez
  J, Stroth U, Qin J and the {TJ-II}~Team 2005 {\em Nuclear Fusion\/} {\bf 45}
  276 \urlprefix\url{http://stacks.iop.org/0029-5515/45/i=4/a=009}

\bibitem{castellano2002well}
Castellano J, Jimenez J~A, Hidalgo C, Pedrosa M~A, Fraguas A~L, Pastor I,
  Herranz J, Alejaldre C and the {TJ-II}~Team 2002 {\em Physics of Plasmas\/}
  {\bf 9} 713--716 \urlprefix\url{http://link.aip.org/link/?PHP/9/713/1}

\bibitem{lopez2009cmode}
L\'opez-Bruna D, Romero J~A, Jim\'enez-G\'omez R, Pedrosa M~A, Ochando M,
  Estrada T, L\'opez-Fraguas A, Medina F, Herranz J, Kalhoff T, Ascas\'ibar E,
  de~la Pe{\~n}a A, Lapayese F and Alonso J 2009 {\em Nuclear Fusion\/} {\bf
  49} 085016 \urlprefix\url{http://stacks.iop.org/0029-5515/49/i=8/a=085016}

\bibitem{castejon2005topology}
Castej\'on F, Fujisawa A, Ida K, Talmadge J~N, Estrada T, L\'opez-Bruna D,
  Hidalgo C, Krupnik L and Melnikov A 2005 {\em Plasma Physics and Controlled
  Fusion\/} {\bf 47} B53
  \urlprefix\url{http://stacks.iop.org/0741-3335/47/i=12B/a=S05}

\bibitem{fernandez2005ebw}
Fern\'andez A, Sarksyan K, Matveev N, Garc\'ia A, Medrano M, Doane J, Moeller
  C, Doncel J, Pardo A, Cappa A, Castej\'on F, Khartchev N, Tereschenko M,
  Tolkachev A and Catal\'an G 2005 {\em Fusion Engineering and Design\/} {\bf
  74} 325 -- 329 ISSN 0920-3796 {Proceedings of the 23rd Symposium of Fusion
  Technology - SOFT 23}
  \urlprefix\url{http://www.sciencedirect.com/science/article/pii/S09203796050%
00384}

\bibitem{garcia2011ebw}
Garc\'ia-Rega{{\~n}}a J~M, Cappa A, Castej\'on F, Caughman J~B~O, Tereshchenko
  M, Ros A, Rasmussen D~A and Wilgen J~B 2011 {\em Plasma Physics and
  Controlled Fusion\/} {\bf 53} 065009
  \urlprefix\url{http://stacks.iop.org/0741-3335/53/i=6/a=065009}

\bibitem{castejon2011ebwbc}
Castej\'on F, Velasco J~L, Garc\'ia-Rega{\~n}a J~M, Allmaier K and Cappa A 2011
  Electron bernstein driven and bootstrap current estimations in the tj-ii
  stellarator {\em Proceedings of the 23rd IAEA Fusion Energy Conference,
  Daejon\/} pp THW/P7--16

\bibitem{hirshman1986dkes}
Hirshman S~P, Shaing K~C, van Rij W~I, Beasley C~O and Crume E~C 1986 {\em
  Physics of Fluids\/} {\bf 29} 2951--2959
  \urlprefix\url{http://link.aip.org/link/?PFL/29/2951/1}

\bibitem{maassberg2005ecd}
Maa{\ss}berg H, Rom\'e M, Erckmann V, Geiger J, Laqua H~P, Marushchenko N~B and
  the W7-AS~Team 2005 {\em Plasma Physics and Controlled Fusion\/} {\bf 47}
  1137 \urlprefix\url{http://stacks.iop.org/0741-3335/47/i=8/a=002}

\bibitem{isaev2009bootstrap}
Isaev M~Y, Watanabe K~Y, Cooper W~A, Yokoyama M, Yamada H, Sauter O, Tran T~M,
  Bergmann A, Beidler C and Maa{\ss}berg H 2009 {\em Nuclear Fusion\/} {\bf 49}
  075013 \urlprefix\url{http://stacks.iop.org/0029-5515/49/i=7/a=075013}

\bibitem{maassberg2009momentum}
Maa{\ss}berg H, Beidler C~D and Turkin Y 2009 {\em Physics of Plasmas\/} {\bf
  16} 072504 \urlprefix\url{http://link.aip.org/link/?PHP/16/072504/1}

\bibitem{nishimura2009viscosity}
Nishimura S, Motojima G, Nakamura Y, Okada H, Kobayashi S, Yamamoto S, Nagasaki
  K, Hanatani K, Kondo K {\em et~al.\/} 2009 {\em Journal of Plasma and Fusion
  Research\/} {\bf 8} 1003
  \urlprefix\url{http://www.jspf.or.jp/JPFRS/PDF/Vol8/jpfrs2009_08-1003.pdf}

\bibitem{beidler2011ICNTS}
Beidler C~D, Allmaier K, Isaev M~Y, Kasilov S~V, Kernbichler W, Leitold G~O,
  Maa{\ss}berg H, Mikkelsen D~R, urakami S~M, Schmidt M, Spong D~A, Tribaldos V
  and Wakasa A 2011 {\em Nuclear Fusion\/} {\bf 51} 076001
  \urlprefix\url{http://stacks.iop.org/0029-5515/51/i=7/a=076001}

\bibitem{ribeiro1987plateau}
Ribeiro E~R~S and Shaing K~C 1987 {\em Physics of Fluids\/} {\bf 30} 462--466
  \urlprefix\url{http://link.aip.org/link/?PFL/30/462/1}

\bibitem{yunta1990bootstrap}
Rodriguez Yunta~A v~R~V~I and P H~S 1990 Bootstrap currents in heliac tj-ii
  configurations {\em Proceedings of the 17th European Conference on Controlled
  Fusion and Plasma Heating, Amsterdam\/} vol 14B p 505

\bibitem{tribaldos2003bootstrap}
Tribaldos~V Maa{\ss}berg~H J~J~A and Varias A 2003 Bootstrap current
  simulations for the tj-ii stellarator {\em Proceedings of the 30th EPS
  Conference on Controlled Fusion and Plasma Physics, St. Petersburg\/} vol 27A
  p P1.28 \urlprefix\url{http://crpppc42.epfl.ch/StPetersburg/PDFS/P1_028.pdf}

\bibitem{allmaier2008variance}
Allmaier K, Kasilov S~V, Kernbichler W and Leitold G~O 2008 {\em Physics of
  Plasmas\/} {\bf 15} 072512
  \urlprefix\url{http://link.aip.org/link/?PHP/15/072512/1}

\bibitem{sugama2002viscosity}
Sugama H and Nishimura S 2002 {\em Physics of Plasmas\/} {\bf 9} 4637--4653
  \urlprefix\url{http://link.aip.org/link/?PHP/9/4637/1}

\bibitem{beidler2007icnts}
Beidler C~D, Isaev M~Y, Kasilov S~V, Kernbichler W, Murakami H~M~S, Nemov V~V,
  Spong D and Tribaldos V 2007 Icnts-impact of incompressible e$\times$ b flow
  in estimating mono-energetic transport coefficients {\em Proceedings of the
  16th Int. Stellarator/Heliotron Workshop, Toki\/} vol NIFS-PROC-69 p P2.O31
  \urlprefix\url{http://www.nifs.ac.jp/itc/itc17/file/PDF_proceedings/poster2/%
P2-031.pdf}

\bibitem{tribaldos2005global}
Tribaldos V and Guasp J 2005 {\em Plasma Physics and Controlled Fusion\/} {\bf
  47} 545 \urlprefix\url{http://stacks.iop.org/0741-3335/47/i=3/a=010}

\bibitem{velasco2009finite}
Velasco J~L, Castejon F and Tarancon A 2009 {\em Physics of Plasmas\/} {\bf 16}
  052303 \urlprefix\url{http://link.aip.org/link/?PHP/16/052303/1}

\bibitem{lotz1988montecarlo}
Lotz W and Nuhrenberg J 1988 {\em Physics of Fluids\/} {\bf 31} 2984--2991
  \urlprefix\url{http://link.aip.org/link/?PFL/31/2984/1}

\bibitem{turkin2011predictive}
Turkin Y, Beidler C~D, Maa{\ss}berg H, Murakami S, Tribaldos V and Wakasa A
  2011 {\em Physics of Plasmas\/} {\bf 18} 022505
  \urlprefix\url{http://link.aip.org/link/?PHP/18/022505/1}

\bibitem{shaing1983neoclassical}
Shaing K~C and Callen J~D 1983 {\em Physics of Fluids\/} {\bf 26} 3315--3326
  \urlprefix\url{http://link.aip.org/link/?PFL/26/3315/1}

\bibitem{nakajima1989bootstrap}
Nakajima N, Okamoto M, Todoroki J, Nakamura Y and Wakatani M 1989 {\em Nuclear
  Fusion\/} {\bf 29} 605
  \urlprefix\url{http://stacks.iop.org/0029-5515/29/i=4/a=006}

\bibitem{kernbichler2002neo}
Kernbichler W, Kasilov S~V, Nemov V~V, Leitold G~O and Heyn M~F 2002 Evaluation
  of the bootstrap current in stellarators {\em Proceedings of the 29th EPS
  Conference on Plasma Physics and Controlled Fusion, Montreux\/} vol 26B p
  P2.100 \urlprefix\url{https://crppwww.epfl.ch/~duval/p2_100.pdf}

\bibitem{spong2005flow}
Spong D~A 2005 {\em Physics of Plasmas\/} {\bf 12} 056114
  \urlprefix\url{http://link.aip.org/link/?PHP/12/056114/1}

\bibitem{estrada2002transient}
Estrada T, de~la Luna E, Ascas\'ibar E, Jim\'enez J~A, Castej\'on F,
  Garc\'ia-Cort\'es I, L\'opez-Fraguas A, S\'anchez J and Tribaldos V 2002 {\em
  Plasma Physics and Controlled Fusion\/} {\bf 44} 1615
  \urlprefix\url{http://stacks.iop.org/0741-3335/44/i=8/a=313}

\bibitem{rapisarda2005rotation}
Rapisarda D, Zurro B, Baciero A, Tribaldos V, Ascas\'ibar E {\em et~al.\/} 2005
  An investigation of the relationship between toroidal rotation and bootstrap
  current in the tj-ii stellarator {\em Europhysics Conference Abstracts: 32nd
  Conf. Plasma Physics, Tarragona\/} vol 29C p P2.086
  \urlprefix\url{http://eps2005.ciemat.es/papers/pdf2/P2_086.pdf}

\bibitem{sanchez2009transitions}
S\'anchez J, Acedo M, Alonso A, Alonso J, Alvarez P, Ascas\'ibar E, Baciero A,
  Balb\'in R, Barrera L, Blanco E, Botija J, de~Bustos A, de~la Cal E, Calvo I,
  Cappa A, Carmona J~M, Carralero D, Carrasco R, Carreras B~A, Castej\'on F,
  Castro R, Catal\'an G, Chmyga A, Chamorro M, Eliseev L, Esteban L, Estrada T,
  Fern\'andez A, Fern\'andez-Gavil\'an R, Ferreira J~A, Fontdecaba J~M, Fuentes
  C, Garc\'ia L, Garc\'ia-Cort\'es I, Garc\'ia-G\'omez R, Garc\'ia-Rega{\~n}a
  J~M, Guasp J, Guimarais L, Happel T, Hernanz J, Herranz J, Hidalgo C,
  Jim\'enez J~A, Jim\'enez-Denche A, Jim\'enez-G\'omez R, Jim\'enez-Rey D,
  Kirpitchev I, Komarov A~D, Kozachok A~S, Krupnik L, Lapayese F, Liniers M,
  L\'opez-Bruna D, L\'opez-Fraguas A, L\'opez-R\'azola J, L\'opez-S\'anchez A,
  Lysenko S, Marcon G, Mart\'in F, Maurin V, McCarthy K~J, Medina F, Medrano M,
  Melnikov A~V, M\'endez P, van Milligen B, Mirones E, Nedzelskiy I~S, Ochando
  M, Olivares J, de~Pablos J~L, Pacios L, Pastor I, Pedrosa M~A, de~la Pe{\~n}a
  A, Pereira A, P\'erez G, P\'erez-Risco D, Petrov A, Petrov S, Portas A,
  Pretty D, Rapisarda D, Ratt\'a G, Reynolds J~M, Rinc\'on E, R\'ios L,
  Rodr\'iguez C, Romero J~A, Ros A, Salas A, S\'anchez M, S\'anchez E,
  S\'anchez-Sarabia E, Sarksian K, Sebasti\'an J~A, Silva C, Schchepetov S,
  Skvortsova N, Solano E~R, Soleto A, Tabar\'es F, Tafalla D, Taranc\'on A,
  Taschev Y, Tera J, Tolkachev A, Tribaldos V, Vargas V~I, Vega J, Velasco G,
  Velasco J~L, Weber M, Wolfers G and Zurro B 2009 {\em Nuclear Fusion\/} {\bf
  49} 104018 \urlprefix\url{http://stacks.iop.org/0029-5515/49/i=10/a=104018}

\bibitem{mynick1983roots}
Mynick H~E and Hitchon W~N~G 1983 {\em Nuclear Fusion\/} {\bf 23} 1053
  \urlprefix\url{http://stacks.iop.org/0029-5515/23/i=8/a=006}

\bibitem{hirshman1986vmec}
Hirshman S~P, van Rij W~I and Merkel P 1986 {\em Computer Physics
  Communications\/} {\bf 43} 143 -- 155 ISSN 0010-4655
  \urlprefix\url{http://www.sciencedirect.com/science/article/pii/001046558690%
0585}

\bibitem{watanabe1992lhd}
Watanabe K, Nakajima N, Okamoto M, Nakamura Y and Wakatani M 1992 {\em Nuclear
  Fusion\/} {\bf 32} 1499
  \urlprefix\url{http://stacks.iop.org/0029-5515/32/i=9/a=I01}

\bibitem{ferrando2004qas}
i~Margalet S~F, Cooper W~A, Allfrey S~J, Popovitch P and Isaev M~Y 2004 {\em
  Fusion Science and Technology\/} {\bf 46} 44
  \urlprefix\url{http://epubs.ans.org/?a=539}

\bibitem{solano1988tj-ii}
Solano E~R, Rome J~A and Hirshman S~P 1988 {\em Nuclear Fusion\/} {\bf 28} 157
  \urlprefix\url{http://stacks.iop.org/0029-5515/28/i=1/a=013}

\end{thebibliography}

\end{document}